\documentclass[12pt]{article}
\usepackage{epsfig}
\usepackage{amsmath}
\usepackage{graphicx}
\usepackage{amssymb}
\def\beq{\begin{equation}}
\def\eeq#1{\label{#1}\end{equation}}
\def\eeqn{\end{equation}}

\def\beqa{\begin{eqnarray}}
\def\eeqa#1{\label{#1}\end{eqnarray}}
\def\eeqan{\end{eqnarray}}

\let\bar=\overbar

\def\Dslash{\not{\hbox{\kern-4pt $D$}}}
\def\dslash{\not{\hbox{\kern-2pt $\del$}}}

\def\msb{{\bar{\ssstyle M \kern -1pt S}}}

\usepackage{fancyhdr,graphicx}
\def\Title#1{\begin{center} {\Large {\bf #1} } \end{center}}

\begin{document}

\Title{Hyperon Stars in Strong Magnetic Fields}

\bigskip\bigskip


\begin{raggedright}

{\it Rosana O. Gomes\index{Gomes, R.O.}, C\'esar A. Zen Vasconcellos \index{Vasconcellos, C.A.Z.}\\
Instituto de F\'isica \\
Universidade Federal do Rio Grande do Sul \\
91501-970 Agronomia\\
Porto Alegre, RS\\
Brazil\\
{\tt Email: r.o.gomes@if.ufrgs.br; cesarzen@if.ufrgs.br}}\\

{\it Veronica Dexheimer \index{Dexheimer, V.A.}\\
Department of Physics \\
Kent State University \\
Kent, OH 44242 \\
USA \\
{\tt Email: vdexheim@kent.edu}}
\bigskip\bigskip
\end{raggedright}

\section{Introduction}
We investigate the effects of strong magnetic fields on the properties of hyperon stars. 
The matter is described by a hadronic model with parametric coupling. The matter is considered to be at zero 
temperature, charge neutral, beta-equilibrated, containing the baryonic octet, electrons and muons. The charged particles have 
their orbital motions Landau-quantized in the presence of strong magnetic fields (SMF). Two parametrisations of a chemical potential dependent 
static magnetic field are considered, reaching $1-2 \times 10^{18}\,G$ in the center of the star. Finally, the Tolman-Oppenheimer-Volkov (TOV) 
equations are solved to obtain the mass-radius relation and population of the stars.  

\section{The Model}

The matter is described in a relativistic mean field formalism. We use a parametric coupling effective model that
considers genuine many-body forces simulated by nonlinear self-couplings interaction terms involving the 
scalar-isoscalar $\sigma$-meson field \cite{taurines2}. The Lagrangian density of our model is defined as:
\begin{equation}
\begin{split}
\mathcal{L}  = 
\sum_{b}\bar{\psi}_b(i\gamma_{\mu}\partial^{\mu} + q_{e}\gamma_{\mu}A^{\mu} -m_b)\psi_b
+\sum_{l}\bar{\psi}_{l}(i\gamma_{\mu}\partial^{\mu} + q_{e}\gamma_{\mu}A^{\mu} -m_{l})\psi_{l} \\
   +\frac{1}{2}(\partial_{\mu}\sigma\partial^{\mu}\sigma-m_{\sigma}^2\sigma^2) 
 +\Big(-\frac{1}{4}\omega_{\mu\nu}\omega^{\mu\nu}+ \frac{1}{2}m_{\omega}^2\omega_{\mu}\omega^{\mu}\Big) 
 +   \Big(-\frac{1}{4} \boldsymbol{\varrho}_{\mu\nu} \cdot
\boldsymbol{\varrho}^{\mu\nu}+ \frac{1}{2}m_{\varrho}
^2\boldsymbol{\varrho}_{\mu} \cdot \boldsymbol{\varrho}^{\mu}\Big) \\ 
-\frac{1}{4}F^{\mu\nu}F_{\mu\nu} 
  + \sum_{b}\big( g^*_{\sigma b}\bar{\psi}_B\psi_b\sigma-g_{\omega
b}\bar{\psi}_B\gamma_{\mu}\psi_b
\omega^{\mu}-\frac{1}{2}g_{\varrho
b}\bar{\psi}_b\gamma_{\mu}\psi_b \boldsymbol{\tau} \cdot
\boldsymbol{\varrho}^{\mu} \big) 
\end{split}\end{equation}
where the hadron-meson coupling is parameterized by:  $g^*_{\sigma b}\equiv \left(1+\frac{g_{\sigma}\sigma}{\lambda m_b}\right)^{-\lambda}g_{\sigma b}$.
The effective mass dependence on $\lambda$ is given by \cite{taurines2}:
\begin{equation}                
m_b^*=m_b-\frac{g_{\sigma b}\sigma_0}{\left(1+\frac{g_{\sigma}\sigma_0 }{\lambda m_b}\right)^{\lambda}}.
\end{equation}

Defining the meson-hyperon couplings as $g_{\eta B} = \chi_{\eta B}\,g_{\eta N}$ for $\eta = \sigma,\omega,\varrho$,
we consider a model based on experimental analysis of hypernucleous data. This model considers that all
hyperon-meson coupling intensities are the same as that of the $\Lambda$-hyperon: 
$\chi_{\sigma B}=\chi_{\sigma\Lambda},\,\chi_{\omega B}=\chi_{\omega\Lambda},\,\chi_{\varrho B}=0$.
The $\Lambda$-hypernucleous binding energy is given by \cite{hys4}: 
\begin{equation}
(B/A)_{\Lambda}=\chi_{\omega B}\,(g_{\omega N}\,\omega_{0}) + \chi_{\sigma B}\,(m_{\Lambda}^{*} -m_{\Lambda}),
\end{equation}
and, in this work, we use $(B/A)_{\Lambda}=-28\, MeV$ at saturation density and $\chi_{\sigma \lambda} = 0.75$.


 

\textbf{Landau Quantization}: The charged particles orbital motion quantization generates the energy spectra:
$E_{\nu, i}=\sqrt{k_{z_i}^{2}+ m^{2}_{i} +2\nu\vert q_{e}\vert{B}},$ where: $m_{i} = m^{*}_{b}$, for baryons and $m_{i} = m_{l}$, for leptons.

We enumerate the Landau levels by $\nu$, which are double degenerated except for the fundamental state.
The largest $\nu$ value for which the $k_{F}^{2}>0$ corresponds to $\quad \nu^{b}_{max} < (\mu_{b}^*)^{2}-(m_{b}^{*})^{2}/2\vert q_{e} \vert{B} \quad$ 
and $\quad \nu^{l}_{max} < (\mu_{l})^{2}-m_{l}^{2}/2\vert q_{e}\vert{B},\quad$  for baryons and leptons, respectively.
Due to interactions, effective chemical potentials for baryons are considered. The chemical potentials are: 
$\quad\mu_{b}^* = E_{F}^{b} + g_{\omega b}\omega_0 + g_{\rho b}\tau_{3 b}\rho_{03}\quad$ for baryons, and  $\mu_l = E_{F}^{l}$ for leptons. 
The fermi energies are: 
$\quad E_{F}^{b} = \sqrt{(k_{F_{b},\nu})^2 + (\overline{m_{b,\nu}})^2}\quad$ and  $\quad E_{F}^{l} = \sqrt{(k_{F_{l},\nu})^2 + (\overline{m_{l,\nu}})^2} $.

The magnetic field modified mass and fermi momentum are defined for baryons and leptons as: 

\begin{equation}
\qquad  \overline{m_{b,\nu}}^2 = (m^{*}_{b})^2 + 2\nu \vert q_{e}\vert{B},\qquad k_{F_{b},\nu}= \sqrt{(\mu^{*}_{b})^2 - (\overline{m_{b,\nu}})^2},
\end{equation}
\begin{equation}  
\qquad \overline{m_{l,\nu}}^2 = m^{2}_{l} + 2\nu \vert q_{e}\vert{B},\qquad k_{F_{l},\nu}= \sqrt{(\mu_{l})^2 - (\overline{m_{l,\nu}})^2}.
\end{equation}

The equation of state of a magnetized neutron star matter ($\varepsilon_{mag}$, $P_{mag}$) was already calculated by \cite{lattimer}. The difference 
in the calculation for our model is that we have a different effective mass expression, which modifies the results for chemical equilibrium.




Considering the pressure isotropy \cite{aurora,mike}, the EoS with all contributions are: 
\\
$\varepsilon = \sum_{B,l} \varepsilon_{mag} + \frac{B^2}{2}, \quad$
$P_{\Vert} = \sum_{b,l} P_{mag} - \frac{B^2}{2}, \quad P_{\bot} = \sum_{b,l} P_{mag} + \frac{B^2}{2} - B\mathcal{M}$.
The magnetization is calculated as: $\mathcal{M}= \partial P_{mag}/\partial B $.




In this work, we use a density dependent magnetic field. We consider a magnetic field with the following baryonic chemical potential dependence \cite{bvariable}:

\begin{equation}
B(\mu)= B_{surf}+B_{c}\left[1-\exp\left(-b\,(\mu_n-938)^{a}\right)\right].
\end{equation}

\begin{figure}[htb]
\begin{minipage}[b]{0.48\linewidth}
\centering
\epsfig{file=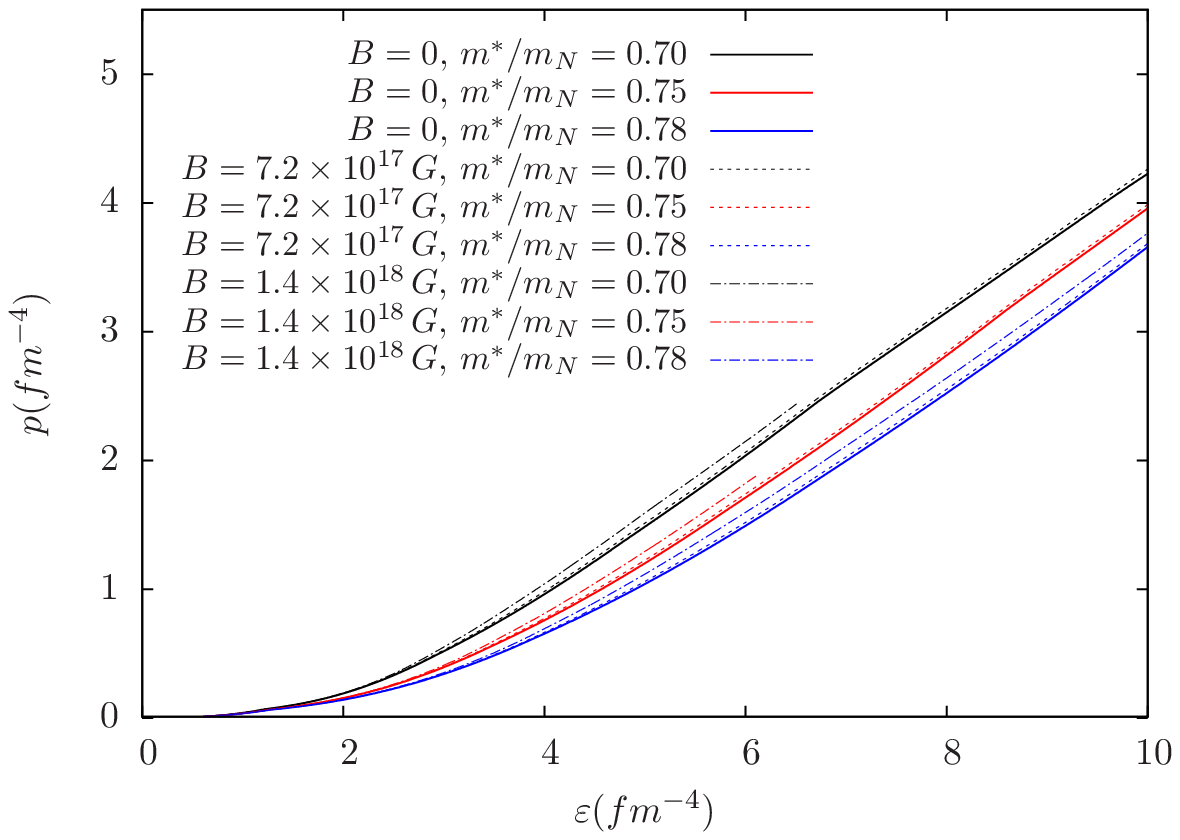,height=2.3in}
\caption{EoS dependence with $B$ for different parameters: $\lambda=0.06;\,0.10;\,0.14$ ($m^*/m_N=0.70;\,0.75;\,0.78\,MeV$)}
\label{eos}
\end{minipage}
\hspace{0.2cm}
\begin{minipage}[b]{0.48\linewidth}
\centering
\epsfig{file=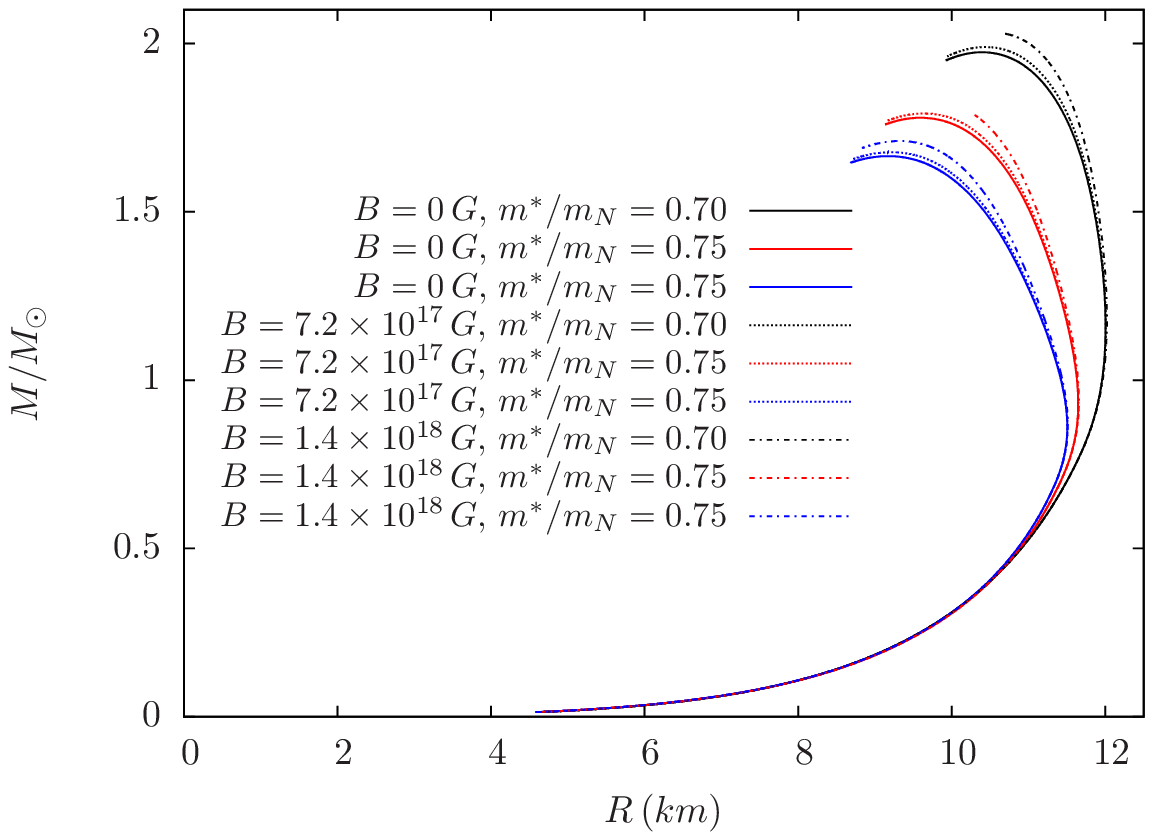,height=2.3in}
\caption{Mass-radius relation for different values of effective mass of the nucleon and central magnetic field.}
\label{tov}
\end{minipage}
\end{figure}

\begin{figure}[htb]
\begin{minipage}[b]{0.48\linewidth}
\centering
\epsfig{file=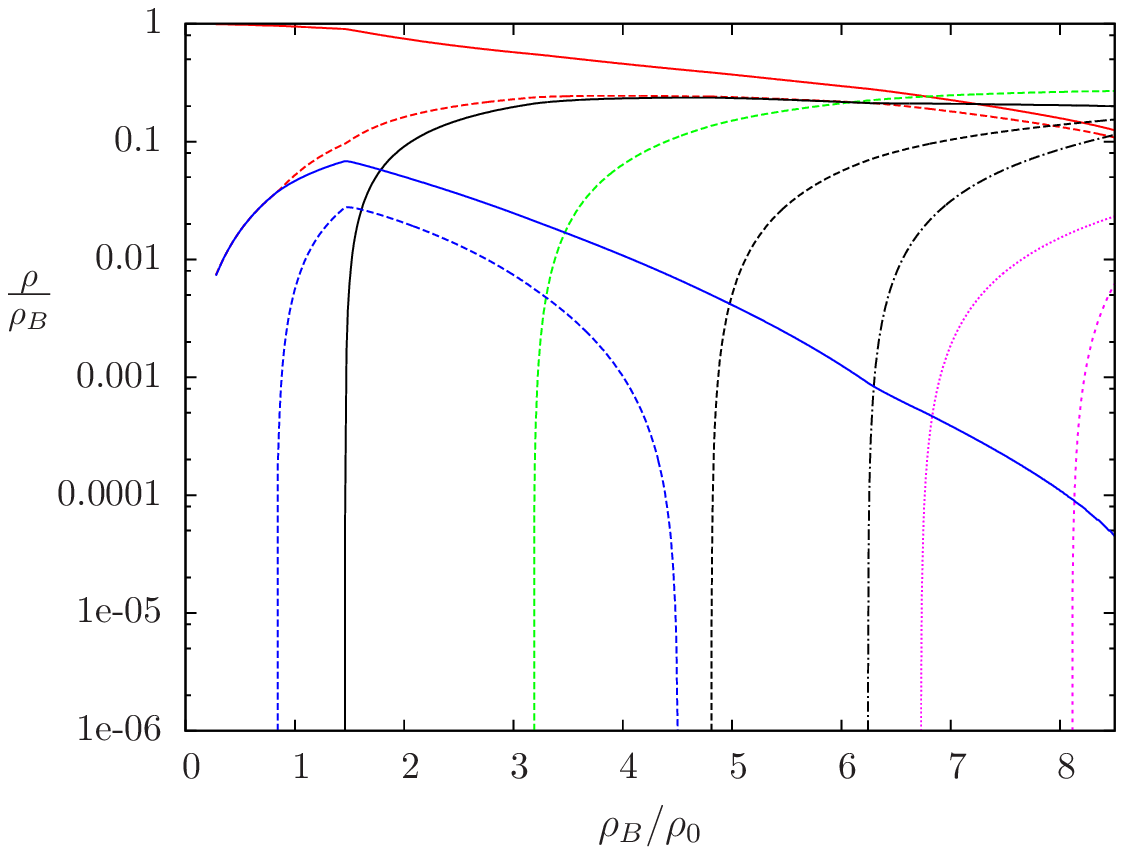,height=2.3in}
\caption{\small{Particles population, $\lambda= 0.06$, $m^*/m_N=0.70$ and $B=0\,G$}.}
\label{popB0}
\end{minipage}
\hspace{0.03cm}
\begin{minipage}[b]{0.48\linewidth}
\centering
\epsfig{file=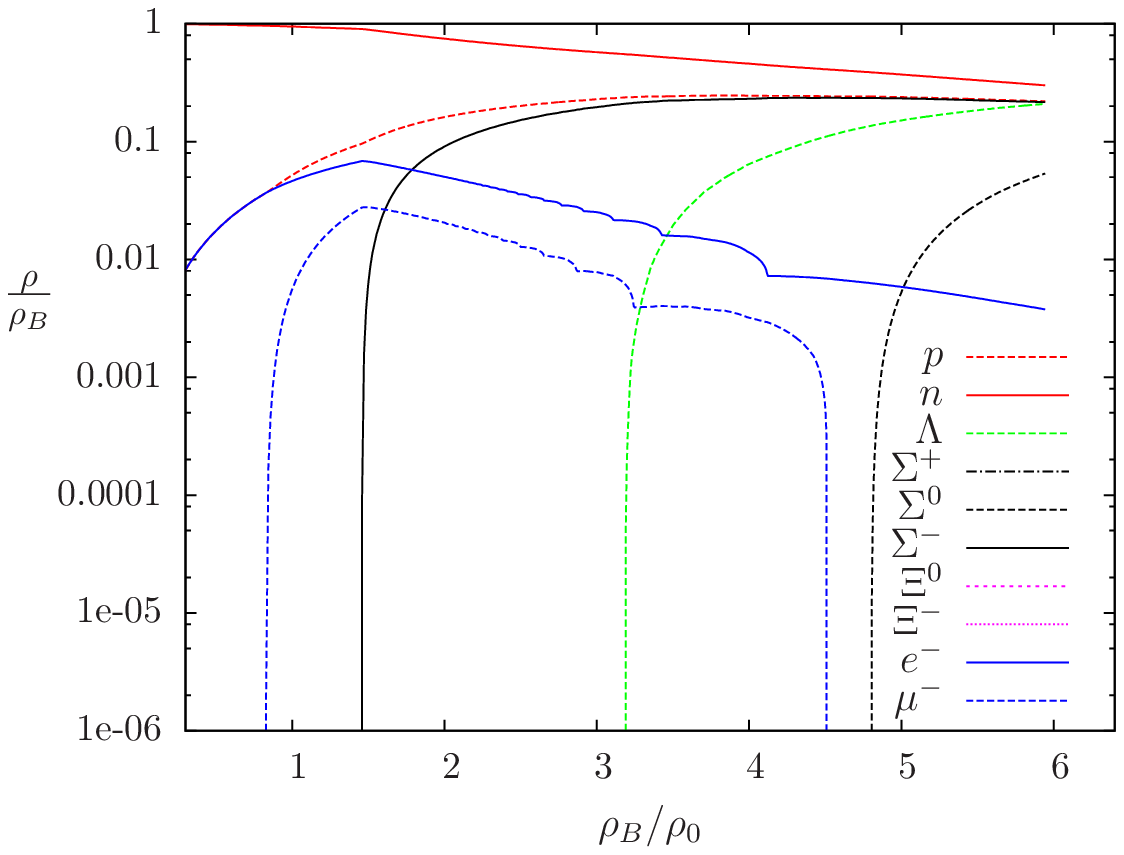,height=2.3in}
\caption{\small{Particles population, $\lambda= 0.06$, $m^*/m_N=0.70$ and $B=4\times10^{18}\,G$}.}
\label{popB}
\end{minipage}
\end{figure}

The $B_{surf}$ and $B_c$ correspond to the magnetic field at the surface of the star, $B_{surf} = 10^{15}\,G$, and at very high baryon chemical potential, 
which we vary. The parameters $a$ and $b$ tell how fast the magnetic field chemical potential dependence is ($a = 2.5;\quad b = 4.35\times 10^{-7}$).



\section{Results and Conclusion}

When all contributions are considered, the total EoS gets stiffer for heigher magnetic fields, Fig~\ref{eos}, due mainly to 
the pure magnetic field contribution. As a consequence, the mass-radius relation permits higher maximum masses for hyperon stars. 
From uncertainties of nuclear matter properties at saturation and our choice of hyperonic coupling scheme, 
the TOV relations \cite{tov} allow us to describe a magnetic hyperon star with $2.03 M_{\odot}$, as shown in Fig~\ref{tov}.
We show in Fig~\ref{popB0},\ref{popB} that the presence of a strong magnetic field supresses
the hyperon population, which is in agreement with \cite{lattimer}.


\begin{thebibliography}{99}

%

\bibitem{taurines2} A. R. Taurines, C. A. Vasconcellos, M. Malheiro, M. Chiapparini, Phys. Rev. C {\bf 63}, 065801 (2001); 
V. A. Dexheimer, C. A. Z. Vasconcellos, B. E. J. Bodmann, Phys. Rev. C {\bf 77}, 065803 (2008)
\bibitem{hys4} N. K. Glendenning and S. A. Moszkowski. Phys. Rev. Lett. {\bf 67}, 2414 (1991); M. Rufa et al., S. A. Moszkowski, Physical Review C {\bf 42}, 2469 (1990); S. Pal, M. Hanauske, I. Zakout, H. Stocker, and W. Greiner, Phys. Rev. C {\bf 60}, 015802 (1999)
\bibitem{bvariable} D. Bandyopadhyay, S. Chakrabarty and S. Pal, Phys. Rev. Lett. {\bf 79}, 2176 (1997); G. Mao, A. Iwamoto and Z. Li, Chin. J. Astron. Astrophys. {\bf Vol. 3}, No. 4, 359-374 (2003),
V. Dexheimer, R. Negreiros, S. Schramm; arxiv: 1108.4479 (2012)
\bibitem{lattimer} A. Broderick, M. Prakash, J.M. Lattimer, ApJ, {\bf 537}, 351. (2000); A. Broderick, M. Prakash, J.M. Lattimer, Physics Letters B, {\bf V. 531}, Issues 3-4, 11, 167 (2002) 
\bibitem{aurora} A. P. Martinez, H. P. Rojas, H. M. Cuesta, Int. J.Mod.Phys. D {\bf 17}, 2107 (2008)
\bibitem{mike} M. Strickland, V. Dexheimer, D. P. Menezes, Phys. Rev. D {\bf 86}, 125032 (2012)
\bibitem{tov} J.R. Oppenheimer, G.M. Volkoff, Phys. Rev. {\bf 55}  347 (1939); R.C. Tolman, Phys. Rev. {\bf 55}  364 (1939)


\end{thebibliography}
\end{document}